
\overfullrule 0pt
\input harvmac

\def\npb{{Nucl.\ Phys.\ }{\bf B}}

\def\plb{{Phys.\ Lett.\ }{\bf B}}

\def\prd{{Phys.\ Rev.\ }{\bf D}}

\def\zpc{Z.\ Phys.\ {\bf C}}


\def \s{\sigma}
\def \b{\beta}
\def \a{\alpha}
\def \g{\gamma}
\def \d{\delta}
\def \e{\epsilon}
\def \l{\lambda}

\def \ph{\phi}

\def \G{\Gamma}

\def \m{\mu}
\def \n{\nu}

\def \r{\rho}

\def\ti{\tilde\imath}
\def\tj{\tilde\jmath}
\def\tk{\tilde k}
\def\tl{\tilde l}
\def\tz{\tilde z}
\def\tA{\tilde A}
\def\tB{\tilde B}
\def\tI{\tilde I}
\def\tJ{\tilde J}
\def\tD{\tilde D}
\def\tF{\tilde F}

\def\pole{{1\over{16\pi^2\e}}}
\def\psib{\overline{\psi}}

\def\Zp{\sqrt{Z_{\ph}}}
\def\bl{\bar \l}
\def\bD{\bar D}

\def\hl{\hat \l}
\def\hB{\hat B}
\def\hD{\hat D}
\def\hF{\hat F}
\def\bF{\bar F}
\def\htD{\hat{\tD}}
\def\htB{\hat{\tB}}
\def\htF{\hat{\tF}}

{\nopagenumbers
\line{\hfil LTH 329}
\vskip .5in
\centerline{\titlefont Equivalence of Dimensional Reduction}
\vskip 5pt
\centerline{\titlefont  and }
\vskip 5pt
 \centerline{\titlefont Dimensional Regularisation}
\vskip 1in
\centerline{\bf I. Jack, D. R. T. Jones and K. L. Roberts}
\bigskip
\centerline{\it DAMTP, University of Liverpool, Liverpool L69 3BX, U.K.}
\vskip .3in
For some years there has been uncertainty over whether
regularisation by dimensional reduction (DRED) is viable for non-supersymmetric
theories. We resolve this issue by showing that DRED is
entirely equivalent to standard dimensional regularisation (DREG),
to all orders in
perturbation theory and for a general renormalisable theory. The two
regularisation schemes
are related by an analytic redefinition of the couplings, under which the
$\b$-functions calculated using DRED transform into
those computed in DREG. The $S$-matrix calculated
using DRED is numerically equal to the DREG version, ensuring that both
schemes give the same physics.
\Date{January 1994}}

\pageno = 1
\newsec{Introduction}
Dimensional reduction (DRED) was introduced some time ago\ref\DRED{W. Siegel,
\plb84 (1979) 193.}
as a means of regulating
supersymmetric gauge theories which maintains manifest supersymmetry whilst
retaining the elegant features of dimensional regularisation (DREG).
However,
its use has often been attended by controversy. Firstly, there are potential
ambiguities in DRED associated with the treatment of the Levi-Civita symbol,
$\e^{\m\n\r\s}$ and, relatedly, with the correct definition of $\g_5$
\ref\eps{H. Nicolai and P. K. Townsend, \plb 93 (1980) 111\semi
W. Siegel, \plb94 (1980) 37\semi
D.~R.~T.~Jones and J.~P.~Leveille, \npb 206 (1982) 473\semi
V. Elias, G. McKeon and R. B. Mann, \npb229 (1983) 487\semi
F. B. Little, R. B. Mann, V. Elias and G. McKeon,  \prd32 (1985) 2707\semi
P. Ensign and K. T. Mahanthappa, \plb194 (1987) 523\semi
J. G. K\"orner, D. Kreimer and K. Schilcher, \zpc54 (1992) 503.}.
We shall not concern ourselves with these problems here.
Secondly, questions have been raised regarding the unitarity of DRED. It was
argued by van Damme and 't Hooft\ref\van{R. van Damme and G. 't Hooft,
\plb150 (1985) 133.}
that whilst DRED preserves unitarity in
the supersymmetric case, unitarity is broken when DRED is applied to
non-supersymmetric theories. However, it is our contention that when DRED is
applied in the fashion envisaged in Ref.~\ref\cjn{D.~M.~Capper, D.~R.~T.~Jones
and P.~van~Nieuwenhuizen, \npb167 (1980) 479.}
 (which differs from that
employed by van Damme and 't Hooft in its treatment of counterterms) then
unitarity is preserved for any theory. In Ref.~\ref\us
{I.~Jack, D.~R.~T.~Jones and
K.~L.~Roberts, ``Dimensional Reduction in Non-supersymmetric Theories'',
Liverpool preprint LTH 320 (Z. Phys. C, to be published).},
we presented evidence for
our claim by computing the two-loop $\b$-functions for a toy model using
both DRED and DREG, and showing that the DRED $\b$-functions could be
transformed into those for DREG by a coupling constant redefinition. We
noted that a regularisation scheme mentioned in Ref.~\van\
(their ``System 3''), which differs from the
version of DRED which they use, led
to the same $\b$-functions as our version of DRED. However it was by no means
clear {\it a priori}
that this scheme was equivalent to the version of DRED which we advocate.
In this paper we shall show that System 3 is indeed equivalent to our
version of DRED, in perfect generality and to all orders.  Since it will also
be clear that System 3 is equivalent to DREG up to coupling constant
redefinition, also to all orders and for a
general theory, we shall have established the exact equivalence of DRED and
DREG and hence we shall have shown that DRED preserves unitarity. Moreover,
we shall also see that the S-matrix computed using DRED is equivalent to
that computed using DREG.

\newsec{Dimensional regularisation and minimal subtraction}
Let us start by describing the standard dimensional regularisation procedure
for a general theory. Consider a general renormalisable gauge theory coupled to
scalar fields (the extension to fermions is straightforward, but introduces
notational complications) with bare Lagrangian
\eqn\eaa{
L(\l_B,g_B,\a_B)=L_G(g_B,\a_B)+\del_{\m}\ph_B^i\del_{\m}\ph_B^i
+\l_B^{ij}\ph_B^i\ph_B^j
+\l_B^{ijk}\ph_B^i\ph_B^j\ph_B^k+\l_B^{ijkl}\ph_B^i\ph_B^j\ph_B^k\ph_B^l  }
in 4 dimensions. In Eq.~\eaa, $\ph_B^i$ represents the bare scalar fields,
and $L_G$
subsumes all the gauge-field dependent terms in the Lagrangian.
The bare gauge coupling is denoted $g_B$
(assuming a simple group so that there is only one gauge
coupling), and $\a_B$ represents the bare gauge-fixing parameter.
The $S$-matrix for the theory is constructed in standard
fashion from the
one-particle-irreducible (1PI) $n$-point functions.
However, it will be sufficient for our purposes to focus our attention
on the 1PI $n$-point functions for $n\le4$, since for a renormalisable theory
in 4 dimensions
these are the only ones which contain potential primitive divergences.
Again, since the theory is renormalisable,
each potentially divergent 1PI $n$-point
function with $n\le4$ is associated with
some $n$-point coupling. For definiteness and notational convenience we shall
restrict our discussion to the 1PI functions with external scalar lines only;
however, the extension to external gauge fields is trivial.
The 1PI 4-point function corresponding to
a 4-point coupling $\l_B^{ijkl}$ will be denoted $\G^{ijkl}$, with a similar
notation for 3- and 2-point functions.
The theory is regulated by continuation to $d=4-\e$
dimensions. We can write, for example,
\eqn\ei{
    \G^{ijkl}=\l^{ijkl}_B+D^{ijkl}(\l_B,g_B,\a_B) }
where
\eqn\ej{
    D^{ijkl}(\l_B,g_B,\a_B)=\sum_{L=1}D^{ijkl}_L(\l_B,g_B,\a_B)  }
with $D^{ijkl}_L(\l_B,g_B,\a_B)$ representing the sum of L-loop diagrams
contributing to $\G^{ijkl}$. (Of course $D^{ijkl}$ and $D^{ijkl}_L$ also
depend on $\e$ and on the external momenta, but we shall not write this
dependence explicitly.)
In terms of $\l_B$ the expression in Eq.~\ei\
has divergences represented by poles in $\e$. Within standard DREG,
we write $\l_B$ as a Laurent series
\eqn\eja{
\l^I_B=\m^{k_I\e}\bigl(\l^I+\sum_{n=1} {{A^I_{n}(\l)}\over{\e^n}}\bigr),
}
(where $\l^I$ represents any of
$\l^{ijkl}$, $\l^{ijk}$, $\l^{ij}$, $g$ or $\a$, and where
$k_I$ takes, for example,
the value 1 when $I$ denotes $ijkl$, $1\over2$ when $I$ denotes $ijk$ and 0
when $I$ denotes $ij$)
so that the Green function $G^{ijkl}$, defined  by
\eqn\ejc{
G^{ijkl}=\Zp^{ii'}\Zp^{jj'}\Zp^{kk'}\Zp^{ll'}\G^{i'j'k'l'}\ ,}
where $Z_{\ph}$ is the wave-function renormalisation for the fields $\ph$,
becomes finite as a function of the renormalised couplings
$\l^I$. Any momentum-dependent terms in the 1PI 4-point and
3-point functions with purely scalar external lines must be finite by
renormalisability, and we
have suppressed them; however this is not so for the 2-point case, for which
we require the Green function
\eqn\ejcb{
G^{ij}=\Zp^{ii'}\Zp^{jj'}\G^{i'j'} ,  }
to be finite, where
\eqn\ejcd{
\G^{ij}=p^2\bigl[\d^{ij}+F^{ij}(\l_B)\bigr]+\bigl[ \l^{ij}_B+
D^{ij}(\l_B)\bigr],}
with $p$ an external momentum.
In fact, Eq.~\ei\ yields a very succinct
formulation of the process of constructing counterterm diagrams, which will be
useful later on in comparing DRED with van Damme and 't Hooft's
System 3.
First of all, in view of Eq.~\ejc\ it is convenient to define ``dressed''
versions of quantities by
\eqn\ejca{
\bl^{ij}=\Zp^{ii'}\Zp^{jj'}\l^{i'j'},  }
and so on.
 Since $G^{ijkl}$ as given by Eqs.~\ejc\ is finite as a function of the
renormalised couplings $\l^I$, we have
\eqn\ejb{
{\rm P.P.}[\bl_B^{ijkl}]=\bl_B^{ijkl}-\l^{ijkl}
=-{\rm P.P.}\{\bD^{ijkl}\} }
where ``P.P.'' stands for ``Pole Part'' and we assume that we are working with
minimal subtraction. (The pole
part is taken only after substituting the expressions Eq.~\eja\ for $\l^I_B$
in terms of the renormalised coupling.)
Next, for convenience, we define $\l_P^I\equiv\sum_{n=1} {{A^I_{n}(\l)}
\over{\e^n}}$,
so that
$\l^I_B=\l^I+\l_P^I$. (Here we set $\m = 1$ for convenience; we know that
$\l^I_P$ contains no explicit $\m$-dependence.)
Expanding $D_L(\l_B)$ around $\l$, $g$ and $\a$, we obtain
\eqn\ek{\eqalign{
\bl_B^{ijkl}&-\l^{ijkl}=\cr
&-{\rm P.P.}\{\sum_{L=1}\sum_{l=1}\Zp^{ii'}\Zp^{jj'}\Zp^{kk'}\Zp^{ll'}
\l_P^{I_1}\ldots\l_P^{I_l}
{d\over {d\l_{I_1}}}\ldots{d\over {d\l_{I_l}}}D^{i'j'k'l'}_L(\l)\},\cr} }
where
\eqn\eka{
\l^I{\del\over{\del\l^I}}\equiv\l^{ijkl}{\del\over{\del\l^{ijkl}}}
+\l^{ijk}{\del\over{\del\l^{ijk}}}+\l^{ij}{\del\over{\del\l^{ij}}}
+g{\del\over{\del g}}+\a{\del\over{\del\a}}.  }
Diagrammatically, each term in the sum on the right-hand side of Eq.~\ek\
represents a set of
diagrams obtained by inserting a
counterterm $\l_P^I$ in turn at each vertex
corresponding to the coupling $\l^I$ in each $L$-loop diagram,
together with an insertion
of $\sqrt{Z_{\ph}}$ on every external line.
Hence the truncation of Eq.~\ek\ at the order in $\l^I$
corresponding to a given loop order $L$, corresponds precisely to
the process of computing counterterms;
the right-hand side of Eq.~\ek\ automatically produces the set of
L-loop diagrams together with the correct associated subtraction diagrams
with the correct symmetry factors.
These are fairly obvious remarks but we have
been unable to find this formulation in the literature.
In a similar fashion, we have from Eqs.~\ejcb\ and
\ejcd\ that
\eqn\ekb{
Z_{\ph}^{ij}=\d^{ij}-{\rm P.P.}\{\bF^{ij}\}. }
Eqs.~\ek\ (with similar expressions for the 3- and 2-point couplings
involving $D^{ijk}$ and $D^{ij}$ respectively)
and \ekb\ clearly determine $\l^I_B$ as Laurent series in
terms of the renormalised couplings as in Eq.~\eja. They also determine
$\sqrt{Z_{\ph}}$ as a Laurent series
\eqn\ekc{
\sqrt{Z_{\ph}}{}^{ij}=\d^{ij}+\sum_{n=1}{z^{ij}_n(\l)\over{\e^n}}.  }
\newsec{Dimensional reduction}
We now turn to discussing two modifications of DREG (which in fact will
turn out to be equivalent); namely DRED, and also van Damme and 't Hooft's
System 3.
The crucial distinction between DRED and DREG is that in DRED the
continuation from 4 to $d=4-\e$ dimensions is made by compactification.
Therefore the number of field components remains unchanged, though they
only depend on $d$ co-ordinates. Consequently $\e$ of the components of
a vector multiplet become scalars, termed ``$\e$-scalars'',
and which we shall denote $\ph^{\ti}$. Moreover,
terms in the Lagrangian involving $\e$-scalars cannot be expected to
renormalise in the same way as terms involving the corresponding vector
fields, since they are not related by $d$-dimensional Lorentz invariance.
Hence we must introduce new couplings for the terms in the Lagrangian
involving $\e$-scalars.
We call these ``evanescent couplings''. We shall term the original fields and
couplings in the Lagrangian ``real''.
In a natural extension of the notation for purely
real couplings in Eq.~\eja, we denote them as $\l_E^{\ti\tj\tk\tl}$,
$\l_E^{\ti\tj kl}$, $\l_E^{\ti jk}$, etc and moreover write them generically as
$\l_E^{\tI}$. It was
demonstrated in Ref.~\us\ that it is vital to maintain the distinction
between evanescent and real couplings in order to show the equivalence
between the DRED and DREG $\b$-functions. The difference between our
implementation of DRED and that adopted by
van Damme and 't Hooft in Ref.~\van\ (their
``System 4'') resides in the counterterms assigned to evanescent couplings.
The renormalisation of evanescent couplings involves computing graphs with
external $\e$-scalars.
Our philosophy is that we should calculate the counterterms required to
make graphs with external $\e$-scalars finite and use those counterterms
to construct subtraction diagrams. This is equivalent to the usual
diagram-by-diagram subtraction procedure where we subtract from each diagram
subtraction diagrams obtained by replacing divergent subdiagrams by
counterterm insertions with the same pole structure. System 4 of
van Damme and 't Hooft,
on the other hand, involves replacing the genuine counterterms for
evanescent couplings by the counterterms for the corresponding real couplings,
and this is their interpretation of DRED.
Henceforth, when we use the term DRED we will refer to the former version of
the
scheme. (We should stress that the two versions of DRED are equivalent for
supersymmetric theories.\ref\jo{I. Jack and H. Osborn, \npb249 (1985)
472.}\van) In either case, the calculations are done using
minimal subtraction, so that the counterterms contain only poles in $\e$.
We must take into account factors of $\e$ from the multiplicity of the
$\e$ scalars so that, for instance, a one-loop subdiagram with a divergent
momentum integral giving an ${\e}^{-1}$ pole but with a factor of $\e$ due to
the presence of $\e$-scalars in the loop would be finite and would
not require a subtraction.

Now let us turn to System 3. We consider first the theory obtained by
performing the reduction to $4-E$ dimensions; this will have a Lagrangian
$L=L(\l_B)+L_E(\l_B,(\l_E)_B)$ where $L_E$ represents the extra terms
involving
the additional scalars (which we shall call $\e$-scalars, although at present
their multiplicity is $E$; shortly we shall set $E=\e$). These terms also
involve bare evanescent couplings
denoted by $(\l_E)_B$. We now write the bare couplings in terms of the
renormalised quantities by continuing to $d=4-\e$ dimensions and
computing the counterterms required to make the theory finite as a
function of the renormalised couplings.
We obtain
\eqn\ea{\eqalign{
\l^I_B&=\m^{k_I\e}\bigl(\l^I+\sum_{n=1}\sum_{m=0}
{{A^I_{nm}(\l,\l_E)E^m}\over{\e^n}}\bigr),\cr
(\l_E^{\tI})_B&=\m^{l_{\tI}\e}\bigl(\l_E^{\tI}+\sum_{n=1}\sum_{m=0}
 {{\tA^{\tI}_{nm}(\l,\l_E)E^m}
\over{\e^n}}\bigr),\cr}
 }
where we assume that the counterterms are defined by minimal subtraction, so
that the $A^I_{nm}$ and $\tA^{\tI}_{nm}$ are independent of $\e$. $k_I$
is as defined following Eq.~\eja, and $l_{\tI}$ in similar fashion.
We also have expressions analogous to Eq.~\ekc\ for the
wave-function renormalisation matrix for the real scalar fields and for the
$\e$-scalars:
\eqn\eaa{\eqalign{
\sqrt{Z_{\ph}}{}^{ij}&=\d^{ij}+\sum_{n=1}\sum_{m=0}
{z^{ij}_{nm}(\l,\l_E)E^m\over{\e^n}},\cr
\sqrt{Z_{\e}}{}^{\ti\tj}&=\d^{\ti\tj}+\sum_{n=1}\sum_{m=0}
{\tz^{\ti\tj}_{nm}(\l,\l_E)E^m\over{\e^n}}.
\cr}}
(There is no wave-function renormalisation matrix mixing real scalars and
$\e$-scalars.)
We now set $E=\e$, so that we have
\eqn\eb{\eqalign{
\l^I_B&=\m^{k_I\e}\bigl(\l^I+\sum_{n=1}\sum_{m=0}
{{A^I_{nm}(\l,\l_E)\e^m}\over{\e^n}}\bigr),
\cr
(\l_E^{\tI})_B&=\m^{l_{\tI}\e}\bigl(\l_E^{\tI}+\sum_{n=1}\sum_{m=0}
{{\tA^{\tI}_{nm}
(\l,\l_E)\e^m}
\over{\e^n}}\bigr), \cr
\sqrt{Z_{\ph}}{}^{ij}&=\d^{ij}+\sum_{n=1}\sum_{m=0}
{z^{ij}_{nm}(\l,\l_E)\e^m\over{\e^n}}, \cr
\sqrt{Z_{\e}}{}^{\ti\tj}&=\d^{\ti\tj}+\sum_{n=1}\sum_{m=0}
{\tz^{\ti\tj}_{nm}(\l,\l_E)\e^m\over{\e^n}}.
\cr }}%
Next we define new couplings $\l'^I$, $\l'^{\tI}_E$ by writing
\eqn\ec{\eqalign{
\l^I_B&=\m^{k_I\e}\bigl(\l'^I+\sum_{n=1}
{{B^I_n(\l',\l'_E)}\over{\e^n}}\bigr)\cr
(\l_E^{\tI})_B&=\m^{l_{\tI}\e}\bigl(\l'^{\tI}_E+\sum_{n=1}
{{\tB^{\tI}_n(\l',\l'_E)}\over{\e^n}}\bigr)\cr}  }
and requiring that there is no explicit $\e$-dependence in $B^I_n$ and
$\tB^{\tI}_n$.
The
couplings $\l'^I$, $\l'^{\tI}_E$ are those of van Damme and 't Hooft's scheme
System 3.
The relation between $\l'$, $\l'_E$ and $\l$, $\l_E$ rapidly becomes very
complex as one looks
beyond one loop; we shall shortly give an explicit expression up to two
loops. It is important to note however that the relation between $\l$, $\l_E$
and $\l'$, $\l'_E$ is {\it analytic}.
Next we define $Z^{\prime ij}_{\ph}(\l',\l'_E,\e)$
and $Z_{\ph}^{\prime\prime kj}(\l',\l'_E,\e)$ by
\eqn\eca{
\sqrt{Z_{\ph}^{\prime\prime}}{}^{ik}(\l',\l'_E,\e)
\sqrt{Z_{\ph}^{\prime}}{}^{kj}(\l',\l'_E,\e)
=\sqrt{Z_{\ph}}{}^{ij}(\l,\l_E,\e)   }
where we require
\eqn\ecb{
\sqrt{Z_{\ph}}{}^{\prime ij}(\l',\l'_E,\e)
=\d^{ij}+\sum{z^{\prime ij}_n(\l',\l'_E)\over{\e^n}}  }
with $z^{\prime ij}$ independent of $\e$, and where $Z''_{\ph}$
is finite as $\e
\rightarrow0$. These requirements specify $Z'_{\ph}$ and $Z''_{\ph}$
uniquely. Similarly, we also define
\eqn\ecc{
\sqrt{Z^{\prime}}{}^{\ti\tj}_{\e}(\l',\l'_E,\e)
=\d^{\ti\tj}+\sum{\tz^{\prime \ti\tj}_n(\l',\l'_E)\over
{\e^n}},  }
together with $Z_{\e}^{\prime\prime\ti\tj}(\l',\l'_E,\e)$,
by $\sqrt{Z''_{\e}}\sqrt{Z'_{\e}}=\sqrt{Z_{\e}}$,
with $Z''_{\e}$ analytic in $\e$ and
$\tz^{\prime \ti\tj}_n$ independent of $\e$. We will later show that
the Laurent expansions of the bare couplings in Eq.~\ec\ are exactly those
obtained using DRED, and that correspondingly $Z'_{\ph}$ and $Z'_{\e}$ are the
real scalar and $\e$-scalar wave-function renormalisations obtained using
DRED.

The $\b$-functions for $\l$, $\l_E$, $\l'$, and $\l'_E$ are defined by
\eqn\ed{\eqalign{
\b^I(\l,\l_E,\e)=\m{d\over {d\m}}\l^I,&\qquad
\b_E^{\tI}(\l,\l_E,\e)=
\m{d\over {d\m}}\l_E^{\tI}  \cr
\b'^I(\l',\l'_E,\e)=\m{d\over {d\m}}\l'^I,&\qquad \b_E'^{\tI}(\l',\l'_E,\e)=
\m{d\over {d\m}}\l'^{\tI}_E\cr}   }
and are given explicitly by
\eqn\ee{\eqalign{
\b^I(\l,\l_E,\e)=-k_I \e\l^I + {\cal D}_I\sum_{m=0}A^I_{1m}\e^m,
&\qquad \b_E^{\tI}(\l,\l_E,\e)=-l_{\tI}\e\l^{\tI}_E +
{\cal D}_{\tI}\sum_{m=0}\tA^{\tI}_{1m}\e^m,
\cr
\b'^I(\l',\l'_E,\e)=-k_I \e\l'^I +{\cal D}'_I B^I_1,
&\qquad \b_E'^{\tI}(\l',\l'_E,\e)=
-l_{\tI}\e\l'^{\tI}_E + {\cal D}'_{\tI}\tB^{\tI}_1\cr  }
}
where
\eqn\ef{
{\cal D}_I = \sum_J k_J\l^J.{\del\over{\del\l^J}} +
\sum_{\tJ} l_{\tJ}\l_E^{\tJ}.{\del\over{\del\l_E^{\tJ}}} - k_I.}

${\cal D}_{\tI}$, ${\cal D}'_I$ and ${\cal D}'_{\tI}$
are defined in similar fashion.
(In fact the effect of the ${\cal D}$ operators  is to multiply an $L$-loop
contribution to $A_{1m}$, $\tA_{1m}$, $B_1$ or $\tB_1$ by $L$.)
By definition,
the $\b$-functions $\b$ and $\b'$
for $\l$ and $\l'$ are related by coupling constant redefinition, and we
have from Eq.~\ed
\eqn\ehd{
\b^{\prime I}(\l', \l'_E, \e )=\bigl(\b.{{\del}\over{\del\l}}+
\b_E.{{\del}\over{\del\l_E}}\bigr)\l^{\prime I}(\l, \l_E, \e), }
or equivalently
\eqn\eg{
\b^I(\l, \l_E, \e)=\bigl(\b'.{{\del}\over{\del\l'}}+
\b'_E.{{\del}\over{\del\l'_E}}\bigr)\l^I(\l', \l'_E, \e).  }
Now the
standard DREG $\b$-function for $\l$, in the absence of the $\e$-scalars, is
clearly given by
\eqn\eha{
\b_{\rm DREG}^I(\l)=
\lim_{\e\rightarrow0}\b^I(\l, \l_E, \e)={\cal D}_I A_{10}^I.}

So if we take the limit as $\e\rightarrow0$ of Eq.~\eg, we obtain, using
 Eqs.~\eha\ that:
\eqn\ega{
\b_{\rm DREG}^I(\l)=\bigl(\b'(\l', \l'_E, 0).{{\del}\over{\del\l'}}+
\b'_E(\l',\l'_E, 0).{{\del}\over{\del\l'_E}}\bigr){\l}^I(\l', \l'_E, 0).  }
Our next step is to show that our prescription for DRED is equivalent to
System 3. In order to do this, we need to consider the way in which the bare
couplings are generated diagrammatically, since our version of DRED
depends crucially on the way in which counterterms are constructed for
divergent subdiagrams. For the theory reduced from 4 to $4-E$ dimensions,
the 1PI function corresponding to a real (but not 2-point) coupling $\l^{I}$
is given by an expression
analogous to Eq.~\ei,
\eqn\ep{
\G^{I}(\l_B, (\l_E)_B)=\l^{I}_B+\sum_{M=0}E^MD^{I}_M(\l_B, (\l_E)_B)}
where $D^{I}_M$ represents the sum of all graphs which produce a multiplicity
factor of $E^M$. The 1PI function
corresponding to an evanescent (but not 2-point)
coupling $\l_E^{\tI}$ is similarly given by
\eqn\epa{
\G^{\tI}(\l_B,(\l_E)_B)=(\l_E^{\tI})_B+\sum_{M=0}E^M\tD^{\tI}_M(\l_B, (\l_E)_B)
 }
The 1PI 2-point function for real fields is given by an expression analogous
to Eq.~\ejcd,
\eqn\epb{
\G^{ij}=p^2[\d^{ij}+\sum_{M=0}E^MF_{M}^{ij}(\l_B, (\l_E)_B)]+[\l^{ij}_B+
\sum_{M=0}E^MD_{M}^{ij}(\l_B, (\l_E)_B)],}
and there is a similar expression for the 1PI 2-point function for
$\e$-scalars.
The definitions of $\l^I_B$ , $(\l^{\tI}_E)_B$ in terms of
$\l^I$ , $\l_E^{\tI}$ in Eq.~\ea, together with the expressions for
$Z_{\ph}^{ij}$ and $Z_{\e}^{\ti\tj}$ in Eq.~\eb,
precisely ensure that the Green functions given by the
``dressed'' versions of the 1PI functions, $G^I\equiv\bar{\G}^I$
and $G^{\tI}\equiv\bar{\G}^{\tI}$, are finite as a function of $\l^I$,
$\l^{\tI}_E$. (Dressed versions of
Green's functions with external $\e$-scalars are
defined in a similar  way to Eq. \ejca, but with factors
of $\sqrt{Z_{\e}}^{\tj\ti}$ instead of $\sqrt{Z_{\phi}}^{ji}$
where appropriate.) To take the example of the real 2-point coupling,
$\sqrt{Z_{\ph}}^{ii'}\sqrt{Z_{\ph}}^{jj'}\G^{i'j'}(\l_B,(\l_E)_B)$
is guaranteed to be
finite. This implies the separate finiteness of
\eqn\epbb{\eqalign{
&\sqrt{Z_{\ph}}^{ii'}(\l,\l_E,\e)\sqrt{Z_{\ph}}^{jj'}(\l,\l_E,\e)\bigl[\d^{i'j'}\cr
&+\sum_{M=0}E^MF_M^{i'j'}(\l^I+\sum_{n=1}\sum_{m=0}
{{A^I_{nm}(\l,\l_E)E^m}\over{\e^n}},\l_E^{\tI}+\sum_{n=1}\sum_{m=0}
{{\tA^{\tI}_{nm}(\l,\l_E)E^m}\over{\e^n}})\bigr] \cr} }
and
\eqn\epc{\eqalign{
&\sqrt{Z_{\ph}}^{ii'}(\l,\l_E,\e)\sqrt{Z_{\ph}}^{jj'}(\l,\l_E,\e)
\bigl[\l_B^{i'j'}\cr
&+\sum_{M=0}E^MD_M^{i'j'}\bigl(\l^I+\sum_{n=1}\sum_{m=0}
{{A^I_{nm}(\l,\l_E)E^m}\over{\e^n}},\l_E^{\tI}+\sum_{n=1}\sum_{m=0}
{{\tA^{\tI}_{nm}(\l,\l_E)E^m}\over{\e^n}})\bigr]. \cr} }

{\it A fortiori}, when we set $E=\e$ in all the Green functions,
we still have finite expressions.
We then rewrite
$\l^I_B=\l^I+\sum_{n=1}\sum_{m=0}
{{A^I_{nm}(\l,\l_E)\e^m}\over{\e^n}}$,
$(\l^{\tI}_E)_B=\l_E^{\tI}+\sum_{n=1}\sum_{m=0}
{{\tA^{\tI}_{nm}(\l,\l_E)E^m}\over{\e^n}}$,
in terms of $\l'^I$, $\l'^{\tI}_E$ according to
Eq.~\ec, and at the same time we rewrite  $Z_{\ph}(\l,\l_E,\e)$
and $Z_{\e}(\l,\l_E,\e)$ as $Z''_{\ph}(\l',\l'_E,\e)Z'_{\ph}(\l',\l'_E,\e)$
and $Z''_{\e}(\l',\l'_E,\e)Z'_{\e}(\l',\l'_E,\e)$ according to
Eqs.~\eca--\ecc, obtaining expressions which are finite as functions of
$\l'$, $\l'_E$. For instance, in the 2-point example above, we now have
that
\eqn\epd{\eqalign{
&\sqrt{Z_{\ph}^{\prime\prime}}{}^{ ii'}(\l',\l'_E,\e)
\sqrt{Z_{\ph}^{\prime}}{}^ {i'i''}(\l',\l'_E,\e)
\sqrt{Z_{\ph}^{\prime\prime}}{}^{ jj'}(\l',\l'_E,\e)
\sqrt{Z_{\ph}^{\prime}}{}^{ j'j''}(\l',\l'_E,\e) \times
\cr&\bigl[\d^{i''j''}+\sum_{M=0}\e^MF_M^{i''j''}\bigl(
\l'^J+\sum {{B^J_n(\l',\l'_E)}\over{\e^n}},
\l'^{\tJ}_E+\sum {{\tB^{\tJ}_n(\l',\l'_E)}
\over{\e^n}}\bigr)\bigr]  \cr}}
and
\eqn\epe{\eqalign{
&\sqrt{Z_{\ph}^{\prime\prime}}{}^{ ii'}(\l',\l'_E,\e)
\sqrt{Z_{\ph}^{\prime}}{}^ {i'i''}(\l',\l'_E,\e)
\sqrt{Z_{\ph}^{\prime\prime}}{}^{ jj'}(\l',\l'_E,\e)
\sqrt{Z_{\ph}^{\prime}}{}^{ j'j''}(\l',\l'_E,\e) \times \cr
&\bigl[\l^{\prime i''j''}+\sum_{M=0}\e^MD_M^{i''j''}\bigl(
\l'^J+\sum {{B^J_n(\l',\l'_E)}\over{\e^n}},
\l'^{\tJ}_E+\sum {{\tB^{\tJ}_n(\l',\l'_E)}
\over{\e^n}}\bigr)\bigr] \cr} }
are finite as functions of $\l'$, $\l'_E$.
We can then remove the factors of $\sqrt{Z''_{\ph}}$ and $\sqrt{Z''_{\e}}$ in
Eqs.~\epd, \epe\ and the other Green functions,
still leaving finite expressions,  since $Z''_{\ph}$ and $Z''_{\e}$ are
analytic in $\e$. These resulting finite quantities can be regarded as
deriving from 1PI functions ``dressed'' by $Z'_{\ph}(\l',\l'_E,\e)$ and
$Z'_{\e}(\l',\l'_E,\e)$. Hence we now have that
\eqn\eq{\eqalign{
\hl'^I+\sum {{\hB^I_n(\l',\l'_E)}\over{\e^n}}&+\sum_{M=0}\e^M\hD^I_M(
\l'^J+\sum {{B^J_n(\l',\l'_E)}\over{\e^n}},
\l'^{\tJ}_E+\sum {{\tB^{\tJ}_n(\l',\l'_E)}
\over{\e^n}})  \cr
\hl'^{\tI}_E+\sum {{\htB{}^{\tI}_n(\l',\l'_E)}\over{\e^n}}&+\sum_{M=0}\e^M
\htD{}^{\tI}_M(\l'^J+\sum {{B^J_n(\l',\l'_E)}\over{\e^n}},
\l'^{\tJ}_E+\sum {{\tB^{\tJ}_n(\l',\l'_E)}
\over{\e^n}}),\cr
Z^{\prime ij}_{\ph}(\l',\l'_E)&+\sum_{M=0}\e^M\hF^{ij}_M(\l'^J+
\sum {{B^J_n(\l',\l'_E)}\over{\e^n}},
\l'^{\tJ}_E+\sum {{\tB^{\tJ}_n(\l',\l'_E)}\over{\e^n}}) ,\cr
Z^{\prime \ti\tj}_{\e}(\l',\l'_E)&+\sum_{M=0}\e^M\htF{}^{\ti\tj}_M(\l'^J+
\sum {{B^J_n(\l',\l'_E)}\over{\e^n}},
\l'^{\tJ}_E+\sum {{\tB^{\tJ}_n(\l',\l'_E)}\over{\e^n}}) \cr
} }
(where for example
$\hl^{ij}\equiv\l^{kl}\sqrt{Z'_{\ph}}^{ik}\sqrt{Z'_{\ph}}^{jl}$)
are finite as a function of $\l'^I$, $\l'^{\tI}_E$.
 The ``hatted'' quantities will later
turn out to be identical to quantities evaluated using DRED and
``dressed'' using the DRED wave-function renormalisations. We therefore have
\eqn\er{\eqalign{
\hl'^I+\sum {{\hB^I_n(\l',\l'_E)}\over{\e^n}}&
=\l'^I\cr&-{\rm P.P.}\Bigl\{\sum_{M=0}\e^M\hD^I_M(
\l'^J+\sum {{B^J_n(\l',\l'_E)}\over{\e^n}},
\l'^{\tJ}_E+\sum {{\tB^{\tJ}_n(\l',\l'_E)}
\over{\e^n}})\Bigr\} ,\cr
\hl^{\prime\tI}_E+\sum {{\htB{}^{\tI}_n(\l',\l'_E)}\over{\e^n}}&=
\l^{\prime\tI}_E
\cr&-{\rm P.P.}\Bigl\{\sum_{M=0}\e^M\htD{}^{\tI}_M
(\l'^J+\sum {{B^J_n(\l',\l'_E)}\over{\e^n}},
\l'^{\tJ}_E+\sum {{\tB^{\tJ}_n(\l',\l'_E)}
\over{\e^n}})\Bigr\},\cr
\sum{{z^{\prime ij}_{n}(\l',\l'_E)}\over{\e^n}}
&=-{\rm P.P.}\Bigl\{ \sum_{M=0}\e^M\hF^{ij}_M(\l'^J+
\sum {{B^J_n(\l',\l'_E)}\over{\e^n}},
\l'^{\tJ}_E+\sum {{\tB^{\tJ}_n(\l',\l'_E)}\over{\e^n}})\Bigr\} ,\cr
\sum{{\tz^{\prime \ti\tj}_{m}(\l',\l'_E)}\over{\e^n}}
&=-{\rm P.P.}\Bigl\{\sum_{M=0}\e^M
\htF{}^{\ti\tj}_M(\l'^J+
\sum {{B^J_n(\l',\l'_E)}\over{\e^n}},
\l'^{\tJ}_E+\sum {{\tB^{\tJ}_n(\l',\l'_E)}\over{\e^n}}) \Bigr\}.\cr}}
By looking in turn at each successive order of $\l$ in Eq.~\er, we may obtain
the
contributions to $B^I_n$, $\tB^{\tI}_n$, $z_n^{\prime ij}$ and
$\tz_n^{\prime\ti\tj}$  at each successive loop order recursively,
starting with the
fact that at one loop $B^I_1=A^I_{10}$ and $B^I_n=0$ for $n\ne1$, and
$\tB^{\tI}_1=A^{\tI}_{10}$ and $\tB^{\tI}_n=0$ for $n\ne1$.
Thus Eq.~\er\
completely
determines $B_n^I$and $\tB_n^{\tI}$.
However, it is clear that the counterterms in DRED are
obtained
recursively by the same process;
the diagrams appearing in DRED are precisely the same as those appearing in
Eq.~\er, the starting point at one loop is the same, and successive pole terms
are determined uniquely by minimal subtraction, corresponding precisely to the
fact that $B^I_n$, $\tB^{\tI}_n$, $z^{\prime ij}$ and $\tz^{\prime\ti\tj}$
in Eq.~\er\ have no dependence on $\e$. Hence we conclude that the expansion
of the bare couplings in terms of renormalised quantities obtained using DRED
is exactly that given in Eq.~\ec, and the wave-function renormalisations for
DRED are those given in Eqs.~\ecb\ and \ecc. In other words, DRED is
identical to System 3. This completes our proof that DRED is equivalent to
DREG up to coupling constant redefinition, since, by virtue of Eq.~\ega,
we already know that System 3 is coupling-constant-redefinition
equivalent to DREG. Although our discussion has been restricted to gauge
theories coupled only to scalar fields, it is straightforward to include
fermions. In fact all our equations remain true if the meaning of $\l^I$
and $\l_E^I$ is extended to include fermion couplings, and if the
``dressing'' by wave-function renormalisations is accomplished by
fermion wave-function renormalisations where appropriate. The fermion
wave-function renormalisations are transformed to System 3 in precisely the
same fashion as the scalar ones.

Our final general result is that the $S$-matrix for DRED is equal to that for
DREG. Let us first discuss processes corresponding to the
various interactions that appear in the Lagrangian.
The contribution to the S-matrix element corresponding to
a given process consists of the appropriate $G^{I}$ divided by
a ${\sqrt{G}}{}^{ij}$ for each external leg.
 For the theory reduced to
$4-E$ dimensions, this can obviously be written $\sum_{M=0}E^MS_M$, where
$S_M$ contain no $E$-dependence. Clearly $S_0$ represents the corresponding
contributions to the S-matrix for DREG. Each term $S_M$ is individually
finite as $\e\rightarrow 0$. Hence, when we set $E=\e$ and let
$\e\rightarrow0$,
we simply obtain the DREG contribution. On the other hand, when we set $E=\e$
and make the redefinitions in Eq.~\ec and Eq.~\eca\ (and of $Z_{\e}$),
we obtain Eq.~\eq\
and when we then let $\e\rightarrow0$ we obtain the DRED
contribution to the $S$-matrix (apart from a finite
wave-function renormalisation on
each external line due to $Z''$).
So the contributions to the $S$-matrix from diagrams corresponding
to $\l^i$ within DRED are precisely equivalent to those within DREG.

So far we have concentrated on Green's functions corresponding to interactions
present in the Lagrangian. The
contributions from diagrams not corresponding to couplings in the Lagrangian,
but with real external particles (i.e. not $\e$-scalars)
are also guaranteed to be the same in the two schemes. In fact it is easy to
see that the formalism deployed in this section demonstrates this also.
The process by which the DREG Green's function is transformed into the
DRED one goes through in the same way, the only difference being that there
is no coupling $\l^{I}$ corresponding to the Green's function $G^I$.
Therefore the complete $S$-matrix
is numerically the same in DRED as in DREG.

\newsec{Discussion}

By this point the reader might be forgiven for wondering why it is
that anyone should wish to use DRED in the non-supersymmetric case. The
most compelling motivation, as far as we are aware, arises in problems
where use of Fierz identities are required, such as, for example, in
the calculation of the anomalous dimensions of four-Fermi operators.
Fierz identities which are straightforward in four dimensions become
problematic for non-integer $d$.
When DRED is used, it is possible to factorise out the Dirac matrix
algebra in a calculation so that it can be performed entirely in
four dimensions\ref\korn{J. G. K\"orner, G. A. Sch\"uler and S. Sakakibara,
\plb194 (1987) 125\semi
J. G. K\"orner and P. Sieben, \npb363 (1991) 65\semi
P. J. O'Donnell and H. K. K. Tung, \prd45 (1992) 4342\semi
P. J. O'Donnell and H. K. K. Tung, Toronto preprint UTPT-91-32\semi
J. G. K\"orner and M. M. Tung, Mainz preprint
MZ-TH/92-41\semi
M.~Misiak, preprint TUM-T31-46/93.} .

On the other hand, however, the
importance of evanescent couplings, which we have emphasised, would appear
at first sight to tend to nullify this advantage. Most actual applications
of DRED have not addressed the evanescent couplings at all, and have
proceeded by implicitly setting them equal to their ``natural'' values.
It is easy now to see that this is in fact perfectly valid. In Section $3$
we established that the DRED and DREG S-matrices for the real
particles are  identical, i.e. that (suppressing all indices)
\eqn\faa{
S(\l) = S(\l', \l'_E)}
where  $\l' = \l'(\l, \l_E)$ and $\l'_E = \l'_E (\l, \l_E)$.

Evidently varying $\l_E$ defines a trajectory in $(\l', \l'_E)$-space without
changing the S-matrix. It follows that we are free to choose a point
on this trajectory such that the $\l'_E$ are indeed equal to their natural
values, for example $h=g$, in the
class of theories considered towards the end of the appendix.
If this is done, however, it should be clear from our analysis that it
would {\it not\/} be possible to relate predictions made at different
values of the renormalisation scale $\mu$ by evolving only the
$\beta$-functions corresponding to the real interactions.

As its advantages become more widely appreciated, we may therefore  expect to
see more widespread adoption of DRED in higher order QCD calculations. In
the case of electroweak processes, the treatment of  $\e^{\m\n\r\s}$
presents special difficulties, as we mentioned in the introduction;
these we will return to elsewhere.
\vskip .5in
\noindent{\bf Acknowledgements}
\vskip 5pt
I.J. and K.L.R. thank the S.E.R.C. for financial support.

\appendix{A}{}
It seems to us worthwhile to show explicitly how $\l'$ and $\l'_E$
are constructed at
low orders in perturbation theory. At one loop, for instance, we have, after
setting $E=\e$ in Eq.~\ea,
\eqn\es{\eqalign{
\l_B^I&=\m^{k_I\e}\bigl(\l^I+A_{11}^{(1)I}+{{A_{10}^{(1)I}}\over{\e}}\bigr),
\cr
(\l_B)_E^I&=\m^{l_{\tI}\e}\bigl(\l_E^{\tI}+\tA_{11}^{(1)\tI}+
{{\tA_{10}^{(1)\tI}}
\over{\e}}\bigr),
\cr }}
where the superscript in brackets denotes the loop order. (It is of course
obvious that $A_{1m}^{(1)I}=\tA_{1m}^{(1)\tI}=0$ for $m>1$.)
Hence it is sufficient to take
\eqn\et{\eqalign{
\l'^I(\l,\l_E)&=\l^I+A_{11}^{(1)I}(\l,\l_E), \qquad B_1^{(1)I}(\l,\l_E)
=A_{10}^{(1)I}(\l,\l_E), \cr
\l_E^{\prime\tI}(\l,\l_E)&=\l_E^{\tI}+\tA_{11}^{(1)\tI}(\l,\l_E), \qquad
\tB_1^{(1)\tI}(\l,\l_E)
=\tA_{10}^{(1)\tI}(\l,\l_E) \cr } }
to obtain a Laurent expansion as in Eq.~\ec\ with the correct properties.
Substituting the expressions for $\l'$ and $\l_E$
in Eq.~\et\ into Eq.~\ehd, we readily
find that the one-loop change in
$\l$ induces a change in the
$\b$-functions
\eqn\eu{
\b^{\prime I}(\l+A_{11}^{(1)}, \l_E+\tA_{11}^{(1)},\e)
=\bigl(\b.{{\del}\over{\del\l}}+
\b_E.{{\del}\over{\del\l_E}}\bigr)(\l^I+A_{11}^{(1)I}), }
which, using $\b^{I(1)}=-k_I\e\l^I+A_{10}^{(1)I}+\e A_{11}^{(1)I}$
and $\b_E^{(1)\tI}=-l_{\tI}\e\l^{\tI}_E+\tA^{(1)\tI}_{10}+\e\tA_{11}^{(1)I}$,
 yields a change in the $\b$-function at the two-loop level given by
\eqn\ev{\eqalign{
\b^{\prime(2)I}(\l,\l_E, \e)-\b^{(2)I}(\l,\l_E, \e)
&=\bigl[A_{10}^{(1)}(\l,\l_E).{{\del}\over{\del\l}}+
\tA_{10}^{(1)}(\l,\l_E).{{\del}\over{\del\l_E}}\bigr]A_{11}^I(\l,\l_E) \cr
&\quad -\bigl[A_{11}^{(1)}(\l,\l_E).{{\del}\over{\del\l}}+
\tA_{11}^{(1)}(\l,\l_E).{{\del}\over{\del\l_E}}\bigr]A_{10}^I(\l,\l_E).\cr} }
This result has the familiar form of a Lie derivative. In the light of our
general results, this should correspond to the difference between the DREG and
DRED $\b$-functions. The supersymmetric case was considered some time ago.
The calculation here was relatively straightforward because in a
supersymmetric theory, if one equates
evanescent couplings to the corresponding real couplings, the
$\b$-functions for the evanescent couplings also become
identical to those for the
corresponding real couplings. This means that one can consistently equate
the evanescent couplings to the real couplings throughout Eq.~\ev, obtaining
\eqn\ew{
\b^{\prime(2)I}(\l,\l,\e)-\b^{(2)I}(\l,\l,\e)
=A_{10}^{(1)}(\l,\l).{{\del}\over{\del\l}}A_{11}^I(\l,\l)
-A_{11}^{(1)}(\l,\l).{{\del}\over{\del\l}}A_{10}^I(\l,\l). }
It was shown in Ref.~\ref\ij{I. Jack, \plb147 (1984) 405.}
 that this difference indeed corresponds to the
difference between the 2-loop DREG and DRED $\b$-functions for a
supersymmetric theory.

However, in the non-supersymmetric case, the $\b$-functions for evanescent
couplings are different from those for the corresponding real couplings even
if one equates the evanescent couplings to the corresponding real ones in
the expressions for the $\b$-functions. It becomes vital to distinguish
evanescent couplings from real ones throughout, and consequently explicit
calculations become somewhat involved. In Ref.~\us\  we explicitly
demonstrated that Eq.~\ev\ indeed gives the correct difference
between the DREG and DRED $\b$-functions for a toy model.

At the next order we have
\eqn\ex{\eqalign{
\l_B^I&=\m^{k_I\e}\bigl(\l^I+A_{11}^{(1)I}+A_{11}^{(2)I}+A_{22}^{(2)I}+
\e A_{12}^{(2)I}\cr  &\quad+{1\over{\e}}[A_{10}^{(1)I}+A_{10}^{(2)I}
+A_{21}^{(2)I}]+{A_{20}^{(2)I}\over{\e^2}}\bigr), \cr
(\l_B)_E^{\tI}&=\m^{l_{\tI}\e}\bigl(\l_E^I+\tA_{11}^{(1)\tI}+
\tA_{11}^{(2)\tI}+\tA_{22}^{(2)\tI}+
\e\tA_{12}^{(2)\tI}\cr
&\quad+{1\over{\e}}[\tA_{10}^{(1)\tI}+\tA_{10}^{(2)\tI}
+\tA_{21}^{(2)\tI}]+{\tA_{20}^{(2)\tI}\over{\e^2}}\bigr). \cr } }
We must now take
\eqn\ey{
\l'^I=\l^I+A_{11}^{(1)I}+A_{11}^{(2)I}+\e A_{12}^{(2)I},}
and
\eqn\ez{\eqalign{
[B_1^{(1)I}(\l',\l'_E)+B_1^{(2)I}(\l',\l'_E)]_{\rm 2-loop}&=
A_{10}^{(2)I}(\l,\l_E)
+A_{21}^{(2)I}(\l,\l_E),\cr
\quad [B_2^{(2)I}(\l',\l'_E)]_{\rm 2-loop}&=A_{20}^{(2)I}(\l,\l_E),\cr}  }
which leads to
\eqn\eza{
B_1^{(2)I}=A_{10}^{(2)I}+A_{21}^{(2)I}-\bigl(A^{(1)}_{11}.{\del\over{\del\l}}
+\tA^{(1)}_{11}.{\del\over{\del\l_E}}\bigr)A_{10}^{(1)I},
\qquad B_2^{(2)I}=A_{20}^{(2)I};  }
there are also entirely analogous results for the evanescent couplings.
Using the consistency condition that relates the double and simple
poles in $\e$, we can simplify $B_1^{(2)I}$ as follows:
\eqn\ezaa{\eqalign{
B_1^{(2)I}&=A_{10}^{(2)I}+
{1\over 2}\bigl[A_{10}^{(1)}.{{\del}\over{\del\l}}+
\tA_{10}^{(1)}.{{\del}\over{\del\l_E}}\bigr]A_{11}^I\cr
&\quad -{1\over 2}\bigl[A_{11}^{(1)}.{{\del}\over{\del\l}}+
\tA_{11}^{(1)}.{{\del}\over{\del\l_E}}\bigr]A_{10}^I.\cr}}
This result is consistent with Eq.~\ev.
We shall not give the full results for the next order; however it is
important to note that terms in $\l'$ proportional to $\e$, as in Eq.~\ey,
play an important role at this and higher orders. At third order, we
write
\eqn\ezb{
\l_B^I=\m^{k_I\e}\bigl(\l'^I+{B_1^I(\l',\l'_E)\over{\e}}
+{B_2^I(\l',\l'_E)\over{\e^2}}+{B_3^I(\l',\l'_E)\over{\e^3}}\bigr).
}
In particular, with $\l'$ as given by Eq.~\ey, we find
$[{B_1^I(\l',\l'_E)\over{\e}}]_{\rm 3-loop}$
contains a finite term $A_{12}^{(2)}.
{\del\over{\del\l}}A_{10}^{(1)I}$, which must be cancelled by a similar term
in $\l'$ at third order.

It is similarly quite easy to derive the forms of the wave-function
renormalisations in Scheme 3 and to compare them with the results obtained
using DRED in particular instances. For example, at one-loop order we have,
after setting $E=\e$ in Eq.~\eaa,
\eqn\ezc{
\sqrt{Z_{\ph}}{}^{ij}=\d^{ij}+z_{11}^{(1)ij}+{z_{10}^{(1)ij}\over{\e}}. }
We find that Eqs.~\eca\ and \ecb\ are satisfied up to one loop by taking
\eqn\ezd{
\sqrt{Z_{\ph}^{\prime\prime}}{}^{ij}=\d^{ij}+z_{11}^{(1)ij}, \qquad
z_1^{\prime (1)ij}=z_{10}^{(1)ij}. }
At two loops we have
\eqn\eze{\eqalign{
\sqrt{Z_{\ph}}{}^{ij}&=\d^{ij}+z_{11}^{(1)ij}+z_{11}^{(2)ij}+
z_{22}^{(2)ij}+\e z_{12}^{(2)ij}\cr
&\quad  +{1\over{\e}}\bigl(z_{10}^{(1)ij}+z_{10}^{(2)ij}+z_{21}^{(2)ij}\bigr)
+{z_{20}^{(2)ij}\over{\e^2}}.\cr}  }
We find, using Eq.~\et, that Eqs.~\eca\ and \ecb\ are satisfied by taking
\eqn\ezf{\eqalign{
\sqrt{Z_{\ph}^{\prime\prime}}{}^{(2)ij}&=
z_{11}^{(2)ij}+z_{22}^{(2)ij}+\e z_{12}^{(2)ij}
-\bigl(A_{11}^{(1)}.{\del\over{\del \l}}+\tA_{11}^{(1)}.{\del\over{\del \l_E}}
\bigr)z_{11}^{(1)ij},\cr
z^{\prime(2)ij}_1&=z_{10}^{(2)ij}+z_{21}^{(2)ij}-
\bigl(A_{11}^{(1)}.{\del\over{\del \l}}+\tA_{11}^{(1)}.{\del\over{\del \l_E}}
\bigr)z_{10}^{(1)ij}-z_{10}^{(1)ik}z_{11}^{(1)kj},\cr
z_2^{\prime(2)ij}&=z_{20}^{(2)ij}.\cr }  }
According to our general results, the primed quantities should
yield the wave-function renormalisation matrix for real scalar fields in
DRED. The most interesting quantity is $z^{\prime(2)ij}_1$,
which is the only one which changes to this order.
We may simplify the
result for $z^{\prime(2)ij}_1$ by using the identity
\eqn\ezg{\eqalign{
z_{21}^{(2)ij}&={1\over2}\bigl[2z_{10}^{(1)ik}z_{11}^{(1)kj}
+\bigl(A_{11}^{(1)}.{\del\over{\del \l}}+\tA_{11}^{(1)}.{\del\over{\del \l_E}}
\bigr)z_{10}^{(1)ij}\cr &\quad
+\bigl(A_{10}^{(1)}.{\del\over{\del \l}}+\tA_{10}^{(1)}.{\del\over{\del \l_E}}
\bigr)z_{11}^{(1)ij}
\bigr] \cr}}
which follows from standard renormalisation group consistency conditions
which determine higher-order poles in terms of simple poles.
We then obtain
from Eqs.~\ezf\ and \ezg\
\eqn\ezh{\eqalign{
z_1^{\prime(2)ij}&=z_{10}^{(2)ij}+
{1\over2}\bigl[
\bigl(A_{10}^{(1)}.{\del\over{\del \l}}+\tA_{10}^{(1)}.{\del\over{\del \l_E}}
\bigr)z_{11}^{(1)ij}\cr &\quad
-\bigl(A_{11}^{(1)}.{\del\over{\del \l}}+\tA_{11}^{(1)}.{\del\over{\del \l_E}}
\bigr)z_{10}^{(1)ij}
\bigr].\cr}}
$z^{\prime(2)ij}_1$ in Eq.~\ezh\ should agree with corresponding quantities
calculated in DRED. (Again, there are similar results for the wave-function
renormalisation matrix for the $\e$-scalars.)

We can easily check Eq.~\ezh\ for  the case of a gauge theory with fermions
but no (real) scalars, as discussed in Section (2) of Ref.~\us, the notation
of which we adopt below. The relevant one-loop renormalisation constants are as
follows:
\eqn\ezk{\eqalign{
Z^{WW} &= Z_{\alpha} =
1 + \pole g^2[({13\over 3} - \alpha - {1\over 3}E)C_2 (G) - {8\over 3}T(R)]\cr
Z^{\psib\psi} &= 1 - \pole [2\alpha g^2  + E h^2] C_2 (R)  \cr
Z_g &= 1 + \pole g^2[{4\over 3}T(R) +({1\over 6}E - {11\over 3})C_2 (G)]\cr
Z_h &= 1 + \pole [(4h^2 - 6 g^2 )C_2 (R) +
2h^2 T(R) - 2h^2 C_2 (G)].
}}
Here $Z^{WW}, Z^{\psib\psi}, Z_{\alpha}$, $Z_g$ and $Z_{h}$
are the renormalisation constants
associated with the gauge field, the fermion multiplet,
the gauge parameter, the gauge coupling and the evanescent Yukawa coupling
respectively.

{}From Eq.~\ezk\ it is easy to show that
$\d Z^{WW} = Z'^{(2)}{}^{WW} - Z^{(2)}{}^{WW}$
is given by
\eqn\ezl{\eqalign{
\d Z^{WW} = &-{1\over 2}{{g^4}\over{(16\pi^2)^2\e}}[
{1\over 6}C_2 (G).2\bigl (({13\over 3} - \alpha)C_2 (G) -
{8\over 3}T(R)\bigr )\cr
&-{\alpha \over 3}C_2 (G).\bigl (-C_2 (G)\bigr ) -
[{4\over 3}T(R)  - {11\over 3}C_2 (G)]\bigl (-{2\over 3}C_2 (G)\bigr )]\cr
&\!\!\!\!\!\! = {{g^4}\over{(16\pi^2)^2\e}}{{1}\over 2}C_2 (G)^2.\cr}}
This result agrees with
the original DRED calculation in Ref.~\cjn\ (see Eq.~(3.12) of that reference).

By a similar calculation it is straightforward to show that
\eqn\ezm{\eqalign{
\d Z^{\psib\psi} &= Z'^{(2)}{}^{\psib\psi} - Z^{(2)}{}^{\psib\psi}\cr
&= {1 \over{(16\pi^2)^2\e}}\biggl[
6g^2h^2 C_2 (R)^2
-h^4\bigl ( 4C_2 (R)^2 - 2 C_2 (R)C_2 (G) +2 C_2 (R) T(R)\bigr )\biggr]\cr}}

Again this agrees with an explicit
calculation of $Z^{(2)}{}^{\psib\psi}$
in the two schemes \ref\unpub{D. R. T.~Jones (unpublished) 1980.}.
\listrefs
\bye